\documentclass[english,twocolumn, superscriptaddress]{revtex4-1}
\usepackage[T1]{fontenc}
\usepackage[latin9]{inputenc}
\usepackage{color}
\usepackage{float}
\usepackage{textcomp}
\usepackage{amsmath}
\usepackage{amssymb}
\usepackage{graphicx}
\usepackage{esint}

\makeatletter
\usepackage{graphicx}

\makeatother

\usepackage{babel}
\begin{document}

\title{Holographic Vector Field Electron Tomography of Three-Dimensional
Nanomagnets}

\author{Daniel Wolf}
\email{Corresponding author (d.wolf@ifw-dresden.de)}

\affiliation{Institute for Solid State Research, IFW Dresden, Helmholtzstr. 20,
01069 Dresden, Germany}

\author{Nicolas Biziere}

\affiliation{CEMES-CNRS 29, rue Jeanne Marvig, B.P. 94347 F-31055, Toulouse Cedex,
France}

\author{Sebastian Sturm}

\affiliation{Institute for Solid State Research, IFW Dresden, Helmholtzstr. 20,
01069 Dresden, Germany}

\author{David Reyes}

\affiliation{CEMES-CNRS 29, rue Jeanne Marvig, B.P. 94347 F-31055, Toulouse Cedex,
France}

\author{Travis Wade}

\affiliation{Laboratoire des Solides Irradies, Ecole Polytechnique, CNRS, CEA,
Universite Paris Saclay, F 91128 Palaiseau, France}

\author{Tore Niermann}

\affiliation{Institute for Optics and Atomic Physics, Technische Universit\"at Berlin,
Strasse des 17. Juni 135, 10623 Berlin, Germany}

\author{Jonas Krehl}

\affiliation{Institute for Solid State Research, IFW Dresden, Helmholtzstr. 20,
01069 Dresden, Germany}

\author{Benedicte Warot-Fonrose}

\affiliation{CEMES-CNRS 29, rue Jeanne Marvig, B.P. 94347 F-31055, Toulouse Cedex,
France}

\author{Bernd B\"uchner}

\affiliation{Institute for Solid State Research, IFW Dresden, Helmholtzstr. 20,
01069 Dresden, Germany}

\author{Etienne Snoeck}

\affiliation{CEMES-CNRS 29, rue Jeanne Marvig, B.P. 94347 F-31055, Toulouse Cedex,
France}

\author{Christophe Gatel}

\affiliation{CEMES-CNRS 29, rue Jeanne Marvig, B.P. 94347 F-31055, Toulouse Cedex,
France}

\author{Axel Lubk}

\affiliation{Institute for Solid State Research, IFW Dresden, Helmholtzstr. 20,
01069 Dresden, Germany}
\begin{abstract}
Complex 3D magnetic textures in nanomagnets exhibit rich physical
properties, for example in their dynamic interaction with external
fields and currents, and play an increasing role for current technological
challenges such as energy-efficient memory devices. To study these
magnetic nanostructures including their dependency on geometry, composition
and crystallinity, a 3D characterization of the magnetic field with
nanometer spatial resolution is indispensable. Here we show how holographic
vector field electron tomography can reconstruct all three components
of magnetic induction as well as the electrostatic potential of a
Co/Cu nanowire with sub 10\,nm spatial resolution. We address the
workflow from acquisition, via image alignment to holographic and
tomographic reconstruction. Combining the obtained tomographic data
with micromagnetic considerations we derive local key magnetic characteristics,
such as magnetization current or exchange stiffness, and demonstrate
how magnetization configurations, such as vortex states in the Co-disks,
depend on small structural variations of the as-grown nanowire.
\end{abstract}
\maketitle

\section*{Introduction}

Novel synthesis methods and the discovery of emerging magnetic phenomena
(e.g., topological non-trivial textures) at the nanoscale triggered
an expansion of nanomagnetism into three dimensions (3D) focusing
on unconventional magnetic configurations with unprecedented properties
\citep{Fernandez-Pacheco2017}. These nanomagnetic configurations
are interesting for prospected \citep{Parkin2008} and realized applications
in field sensing \citep{karnaushenko_self-assembled_2015}, magnetic
memory, and spintronics \citep{Parkin2008,hoffmann_opportunities_2015,fruchart_bloch-point-mediated_2018}.
Indeed, even trivial magnetic systems such as homogeneously magnetized
domains adopt a 3D modulation near surfaces due to ubiquitous surface
anisotropies and the role of dipolar interactions as the demagnetizing
field. Surface induced modulations play a crucial role in the formation
and the dynamics of different classes of domain walls in magnetic
nanowires (transverse-vortex and Bloch point domain walls \citep{Biziere2013,fruchart_bloch-point-mediated_2018}).
Similarly, surfaces of chiral magnets characterized by a non-zero
antisymmetric Dzyaloshinskii-Moriya interaction (DMI) \citep{Bogdanov1989}
exhibit 3D surface twists \citep{Rybakov2013}, chiral bobbers \citep{Rybakov2016,Zheng2018}
and other 3D textures. Conceptually similar to DMIs, handed magnetic
coupling can occur in curvilinear surfaces, which can be considered
as a frozen and frustrated gauge background induced by the spatially
varying frame of reference. As a consequence, a family of curvature-driven
effects are predicted including magnetochiral effects and topologically
induced magnetization patterning, e.g., skyrmion multiplet states
in bumpy films \citep{Kravchuk2018}. Another example pertains to
magnetically frustrated 3D configurations found in (artificial) spin
ice \citep{Nisoli2013}, where the magnetic dipole interaction is
the driving force for highly degenerated ground states.

To gain insight into 3D magnetic textures in nanomagnets, tomographic
magnetization mapping techniques with spatial resolution approaching
the nanometer regime are highly desired. There are, however, a limited
number of contrast mechanisms, which allow to record projections of
the magnetic field suitable for a tomographic reconstruction, notably,
(soft) X-ray techniques and transmission electron microscopy (TEM)-based
phase reconstruction techniques. Both techniques are still in their
infancy with only a few studies reported so far. Soft X-ray tomography
has been employed to retrieve spin textures \citep{Streubel2015}
on curved magnetic films at a spatial resolution approaching 50\,nm,
with challenges facing the further reduction of resolution and the
limited penetration depth. This could recently be solved by employing
hard X-rays to tomographically reconstruct either one magnetization
component \citep{suzuki_three-dimensional_2018}, or two via a dual-axis
approach, enabling estimation of the third \citep{donnelly_tomographic_2018}.

The first TEM-based proof-of-principle 3D reconstruction of one Cartesian
component of the magnetic induction (flux density) $\mathbf{B}$ by
employing electron holographic tomography (EHT) dates back already
25 years \citep{Lai1994}. However, due to instrumental and computational
limitations, progress in this research field was only little until
five years ago, when a quantitative reconstruction of one $\mathbf{B}$-component
in- and outside magnetic nanoscale materials was achieved \citep{Lubk(2014)d,Wolf2015,Simon2016}.
The spatial resolution obtained in these studies is better than ten
nanometers. However, further improvements in resolution will be very
challenging because of fundamental detection limits for electron holography
(EH) of the magnetic signal \citep{Lichte2008a,Roder(2014)a} and
experimental challenges in electron tomography. The full reconstruction
of all three Cartesian components, which we here refer to as holographic
vector field electron tomography (VFET), remains challenging due to
the complexity of acquisition and reconstruction involved. Notwithstanding,
isolated studies have been reported with large regularization and
reduced spatial resolution \citep{Phatak(2010)} or angular sampling
\citep{Tanigaki(2015)}. Another challenge is to characterize the
specimen in terms of magnetization $\mathbf{M}$ instead of the $\mathbf{B}$-field
reconstructed by EHT. Therefore micromagnetic modelling has been included
in the tomographic reconstruction to retrieve the full magnetic configuration
\citep{Caron2018,Mohan2018}. 

In off-axis electron holography, the phase shift $\varphi$ in the
object plane $\left(x,y\right)$ with respect to a vacuum reference
acquired by an electron wave passing through a magnetic sample along
$z$ direction is given by the Aharonov-Bohm phase shift \citep{Aharonov1959}
\begin{equation}
\varphi\left(x,y\right)=\intop_{-\infty}^{+\infty}\left(\frac{e}{\hbar v}V\left(x,y,z\right)-\frac{e}{\hbar}A_{z}\left(x,y,z\right)\right)\mathbf{\mathrm{d}}z.
\end{equation}
Here $v$ is the electron velocity, $V\left(x,y,z\right)$ the 3D
electrostatic potential, $e$ the elementary charge, $\hbar$ the
reduced Planck constant, and $A_{z}\left(x,y,z\right)$ the 3D component
of the magnetic vector potential parallel to the electron beam direction
$z$. Accordingly, the first term can be considered as electric phase
shift $\varphi_{\text{el}}$ and the second one as magnetic phase
shift $\varphi_{\text{mag}}$. By converting the latter to the magnetic
flux enclosed between interfering paths of object and reference wave,
the directional spatial derivatives of the phase are
\begin{eqnarray}
\mathrm{\nabla}_{x,y}\varphi_{\mathrm{mag}}\left(x,y\right) & = & \frac{e}{\hbar}\mathrm{\nabla}_{x,y}\iint\mathbf{B}\left(\mathbf{r}\right)\mathrm{d}\mathbf{S}\nonumber \\
 & = & \frac{e}{\hbar}\int_{-\infty}^{\infty}\begin{pmatrix}B_{y}\left(\mathbf{r}\right)\\
-B_{x}\left(\mathbf{r}\right)
\end{pmatrix}\mathrm{d}z\,,\label{eq:PhaseGrad_vs_B}
\end{eqnarray}
proportional to the projected in-plane $\mathbf{B}$-components. The
last line of the above relation also holds for inline holographic
techniques, such as transport of intensity holography \citep{Teague(1983),Phatak(2010),Lubk2018a}.
Inline techniques, however, act spatially like high-pass filters \citep{Lubk2018a},
i.e., are ill-conditioned for reconstructing low spatial frequencies.
To separate electric and magnetic contributions in the total phase
shift ($\varphi=\varphi_{\mathrm{mag}}+\varphi_{\mathrm{el}}$), two
phase images with reversed magnetic induction are required \citep{Kasama2011}.
Subtracting both allows to remove the electric phase shift, which
does not change sign under time-reversal as opposed to the magnetic
one. To reconstruct the 3D distribution of $\mathbf{B}$ from their
projections, we collect a series of projections at different angles
(tilt series) and employ tomographic reconstruction algorithms to
this tilt series. By tilting about a particular axis, say $x$, we
obtain a complete set of projections of the component $B_{x}$, whereas
the remaining two components mix in their contribution to the magnetic
flux as a function of tilt angle. Consequently, three tilt series
around perpendicular axes are required to reconstruct the vector field
$\mathbf{B}$. Commercially available TEM specimen holder allow tilting
only about two independent axes. Therefore, we have to exploit the
solenoidal character of the \textbf{$\mathbf{B}$}-field (Gauss's
law for magnetism), i.e. 
\begin{equation}
\mathbf{\nabla}\cdot\mathbf{B}=\partial_{x}B_{x}+\partial_{y}B_{y}+\partial_{z}B_{z}=0\thinspace,\label{eq:Gauss's law for magnetism}
\end{equation}
to obtain the third component $B_{z}$ (see Supplementary Note~1).

\textcolor{black}{In the following, we combine all these strands,
facilitating the 3D reconstruction of the magnetic induction as well
as the 3D magnetization} \textcolor{black}{and related magnetic properties.}
More specifically, we demonstrate the 3D reconstruction of all three
$\mathbf{B}$-field components with sub $10\,\mathrm{nm}$ resolution
and derive magnetization currents and exchange energy contributions.
We correlate the magnetic configuration to the chemical composition
from the mean inner potential (MIP) 3D distribution reconstructed
in parallel by holographic VFET. Additionally, we show how to derive
$\mathbf{\boldsymbol{M}}$ from $\boldsymbol{\mathbf{B}}$ by involving
micromagnetic considerations. Within the case study of a multilayered
nanowire (NW) composed of alternating magnetic Co and non-magnetic
Cu disks, we observe a range of different magnetic configurations
in\textcolor{black}{cluding vortex states. Such structures are model
systems of spintronics devices such as spin valves or spin-torque-based
microwave devices, for which the knowledge and control of the initial
magnetic state is crucial for applications.}

\section*{Results}

\subsection*{Workflow of holographic vector field electron tomography}

\begin{figure}[H]
\includegraphics[width=1\columnwidth]{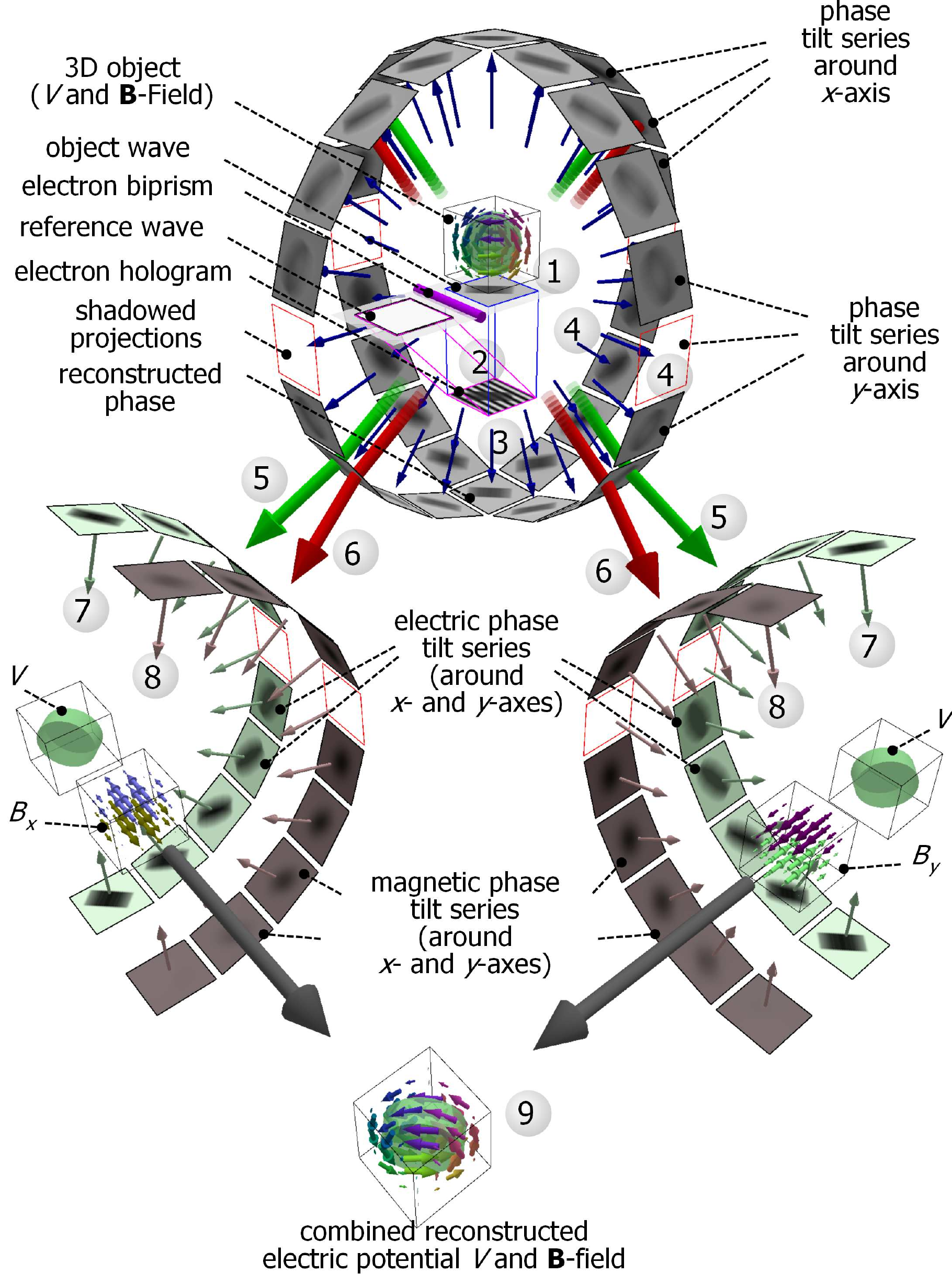}\caption{\label{fig:Principle-of-EHVT}Principle of holographic vector field
electron tomography. 1) Electron waves pass through the magnetic sample
represented by both electrostatic potential $V$ and magnetic $\mathbf{B}$-field.
2) Underneath the sample, an electrostatic biprism brings the modulated
part (object wave) of the electron waves to interference with an unmodulated
part (reference wave) resulting in an electron hologram at the detector
plane. 3) The hologram is reconstructed numerically yielding the amplitude
and phase shift of the object wave. 4) Rotating and recording the
object around two tilt axes, $x$ and $y$, provides for each tilt
axis a $360{^\circ}$ phase tilt series. A few projections might be missing
due to technical and space limitations of the experiment. 5/6) Half
the sum/difference of opposite projections (green/red arrow) results
in the electric/magnetic part of the phase shift, thus two $180{^\circ}$ tilt
series for the tilt axes $x$ and $y$. 7) The electric potential
is reconstructed in 3D from the electric phase tilt series around
$x$- or $y$-axis, or both results are averaged. 8) The two magnetic
$\mathbf{B}$-field components $B_{x}$ and $B_{y}$ are reconstructed
in 3D from the magnetic phase tilt series around $x$- and $y$-axis,
respectively. 9) Finally, the 3D $B_{z}$component is obtained by
solving $\nabla\cdot\boldsymbol{\mathbf{B}}=0$. }
\end{figure}

Holographic VFET, illustrated in Fig.~\ref{fig:Principle-of-EHVT},
starts with the electron hologram acquisition (1, 2) and ensuing phase
image reconstruction out of it (3). Then, these three steps are repeated
at each tilt direction for the collection of two orthogonal tilt series
of phase images (4). The latter two are separated in their electric
(5) and magnetic (6) parts, and reconstructed by tomographic techniques
yielding the 3D distributions of electric potential (7) as well as
the two magnetic $\mathbf{B}$-field components $B_{x}$ and $B_{y}$
(8). Finally, the third magnetic $\mathbf{B}$-field component $B_{z}$
is computed from $\nabla\cdot\boldsymbol{\mathbf{B}}=0$ (9). In order
to perform the crucial separation of electric and magnetic phase shifts
(5,6), each tilt series (ideally  $360{^\circ}$ is split into two sub tilt
series (ideally $180{^\circ}$). However, even by using dedicated tomography
TEM sample holders, in many cases the specimen and holder geometry
may de facto limit the tilt range to usually $140{^\circ}$. Thus,
the second sub tilt series with the corresponding opposite projections
has to be recorded after the sample was turned upside-down outside
the electron microscope. Consequently, tomograms reconstructed from
those tilt series with incomplete tilt range suffer from a direction-dependent
reduction of resolution leading to so-called missing wedge artifacts
\citep{Midgley2003}. The details about acquisition, holographic reconstruction,
the elaborate post processing of projection data, tomographic reconstruction,
and computation of the $B_{z}$-component are given in the Methods
section at the end.

\subsection*{3D magnetic induction mapping of a layered Cu/Co nanowire}

\begin{figure*}
\includegraphics[width=0.9\textwidth]{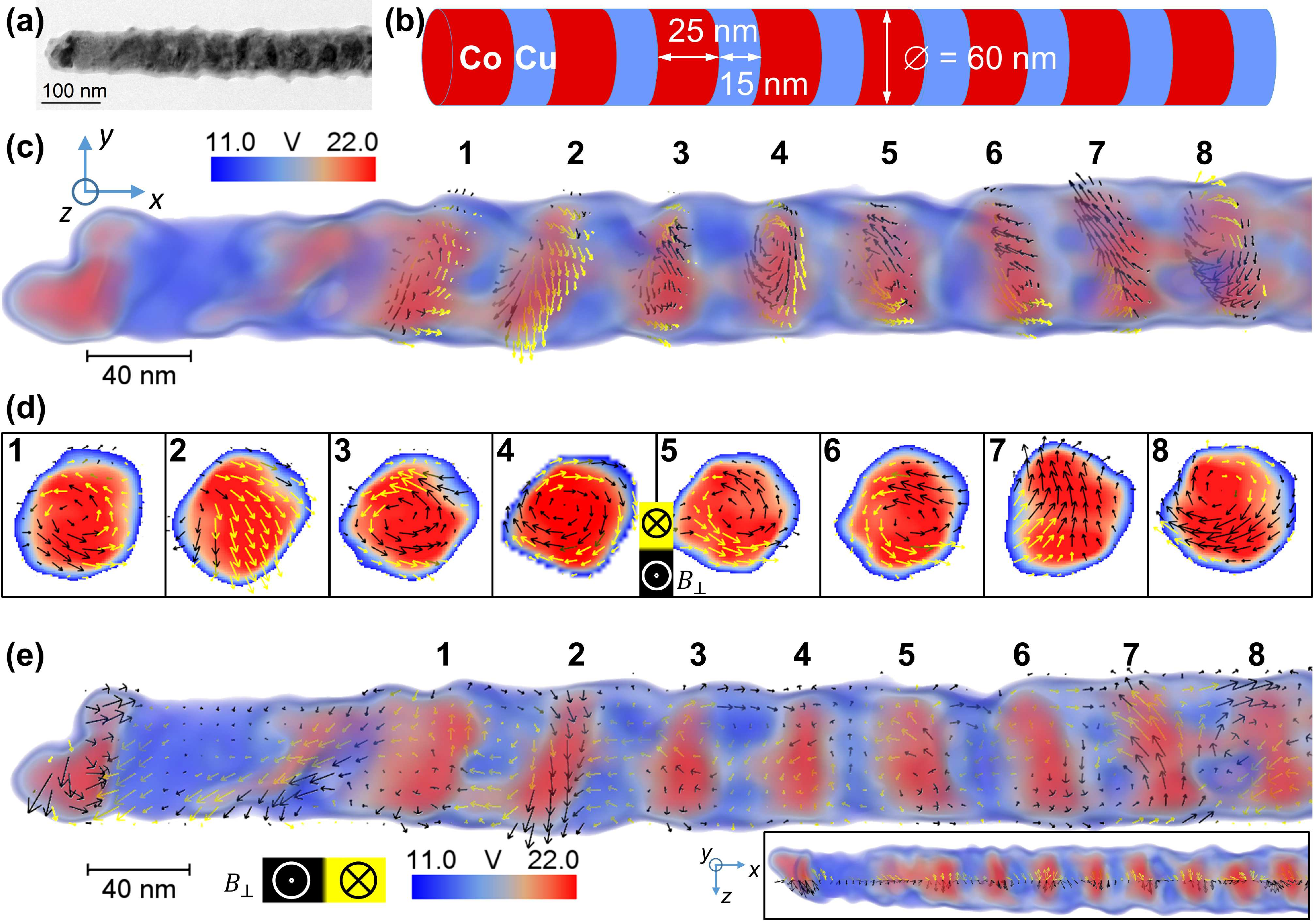}

\caption{3D reconstruction of multilayered Co/Cu nanowire (NW). (a) Bright-field
transmission electron micrograph of the NW. (b) Idealized model of
the NW as intended by template-based electrodeposition growth. (c)
The volume rendering of the NW's mean inner potential (MIP) displays
Co in red and Cu in bluish-white (as illustrated in (b)). The arrow
plots represent the field lines of magnetic induction ($\mathbf{B}$-field)
sliced through eight Co disks, where the direction of the normal $\mathbf{B}$-field
component is color-coded in black and yellow. \textcolor{black}{Please
note that the coordinate system is rotated with respect to the experimental
setup: The $B_{x}$ component is now parallel to the NW axis (=out-of-plane
component of disks).} (d) Slices through the Co disks as indicated
in (c): Co disks 2,7 show in-plane magnetization, 1,3,5,6 counter-clock-wise,
and 4,8 clock-wise vortex magnetization. (e) Same viewing direction
as (c), but arrow plot taken in axial NW-direction. The inset illustrates
where the arrow plot intersects the NW.\label{fig:3D-reconstruction}}
\end{figure*}

Fig.~\ref{fig:3D-reconstruction} depicts the 3D reconstruction of
a multilayered Co/Cu NW using holographic VFET. The bright-field TEM
image shown in Fig.~\ref{fig:3D-reconstruction}a reveals the nanocrystalline
structure of the NW, but does not visualize the alternating Co and
Cu segments as intended by template-based electrodeposition growth
(Fig.~\ref{fig:3D-reconstruction}b). Fig.~\ref{fig:3D-reconstruction}c
shows the 3D $\mathbf{B}$-field reconstructed from the magnetic phase
shift in combination with the MIP obtained from the electric phase
shift distribution reflecting the composition of the nanowire. The
positions of the stacked Co and Cu disks (visible due to the difference
in the MIP between Co and Cu) are also confirmed by energy-filtered
TEM (EFTEM) on exactly the same NW (see Supplementary Note\,2). Previous
quantitative EFTEM investigations on similar NWs have also demonstrated
the presence of 15\% of Cu in the Co disks resulting from the electrodeposition
\citep{Reyes2016}. We obtained MIP values for the Co (incl. 15\%
Cu) and Cu disks of $V_{0}^{\text{Co}}=\left(21\pm1\right)\text{V}$
and $V_{0}^{\text{Cu}}=\left(17.5\pm1\right)\text{V}$, respectively
by determining the peak maximum and width in the histograms evaluated
at the corresponding tomogram regions. The MIP tomogram (Fig.~\ref{fig:3D-reconstruction}c
and Supplementary Movie~1) already provides an important contribution
for a better analysis of magnetic properties, because it reveals that
the Co disks not only deviate slightly from their nominal thickness
of $25\,\text{nm}$ and their cylindrical shape intended by electrodeposition
synthesis (see Methods section for the details), but also vary to
some degree in the inclination angle along the growth direction. Therefore,
to illustrate the 3D distribution of the $\mathbf{B}$-field within
the individual Co disks, we selected cross-sections oriented parallel
to their inclined base surface (Figs. \ref{fig:3D-reconstruction}c,d
and Supplementary Movie~2). Accordingly, two different magnetic configurations
were observed: A homogeneously in-plane magnetized state and a vortex
state. The latter shows both clock-wise and counter-clock-wise rotation
without noticeable correlation of the rotation between the Co-disks
(i.e., coupling between different rotational directions). At the center
of the vortex, the magnetization (and hence the $\mathbf{B}$-field)
is expected to rotate out-of-plane within a core radius smaller than
$10\,\text{nm}$ (see further below for details), which is, however,
difficult to resolve unambiguously in the vector tomogram at the present
spatial and signal resolution. These out-of-plane magnetized vortex
cores together with other out-of-plane components, such as Co bridges
in the Cu (e.g., between disks 1 and 2, 7 and 8 in Fig. \ref{fig:3D-reconstruction}e)
and asymmetries of the vortices with respect to the NW axis, may,
however, contribute to an Aharonov-Bohm phase shift in the vacuum
region around the NW. This is demonstrated in Supplementary Note~3,
where the reconstructed 3D magnetic structure is correlated with a
corresponding phase image: An inspection of the vacuum region in the
phase image reveals that the core polarities are mutually aligned
by their long-range dipole interaction (producing a net flux density
along the NW), while the in-plane states (disks 2 and 7) emanate particularly
strong stray fields. We finally observe some out-of-plane modulations
outside of the core region as indicated by yellow and black color
in (Fig. \ref{fig:3D-reconstruction}d), which will be discussed in
detail further below. Turning to the longitudinal cross section (Fig.~\ref{fig:3D-reconstruction}e)
we notice the strong reduction of the magnetic induction between the
vortex state discs (producing small stray fields only). Correspondingly,
significant $\mathbf{B}$-fields are visible in the vicinity of the
in-plane magnetized discs producing large stray fields. The longitudinal
section also exhibits the out-of-plane modulations of the vortex states
and indicates a complicated configuration in the NW tip, which will
not be discussed further in detail. In order to examine the reliability
and fidelity of these delicate findings, we measured the spatial resolution
of the tomograms by Fourier shell correlation (FSC) \citep{Heel2005}.
As described in Supplementary Note~4, we determined with FSC a spatial
resolution of about 7\,nm for the 3D reconstructions of both $B_{x}$
and $B_{y}$, and of about $5\,\text{nm}$ in case of the 3D electrostatic
potential. In addition, we verified the fidelity of the reconstructed
$\mathbf{B}$-field by applying our holographic VFET reconstruction
on a simulated magnetic Co disk with similar magnetization, dimension,
orientation, configuration (vortex and in-plane magnetized) and sampling
as in the experiment (Supplementary Note~5). The resulting $\mathbf{B}$-field
tomograms agree very well with their simulated input data, even though
a limited tilt range is used ($\pm70{^\circ}$). Our VFET analysis
can thus unambiguously reveal the 3D magnetic structure of a complex
system, highlights the various magnetic configurations and determine
magnetic characteristics such as the vorticity and out-of-plane modulations
of a vortex.

\subsection*{Magnetization current exchange density}

\begin{figure}[h]
\includegraphics[width=1\columnwidth]{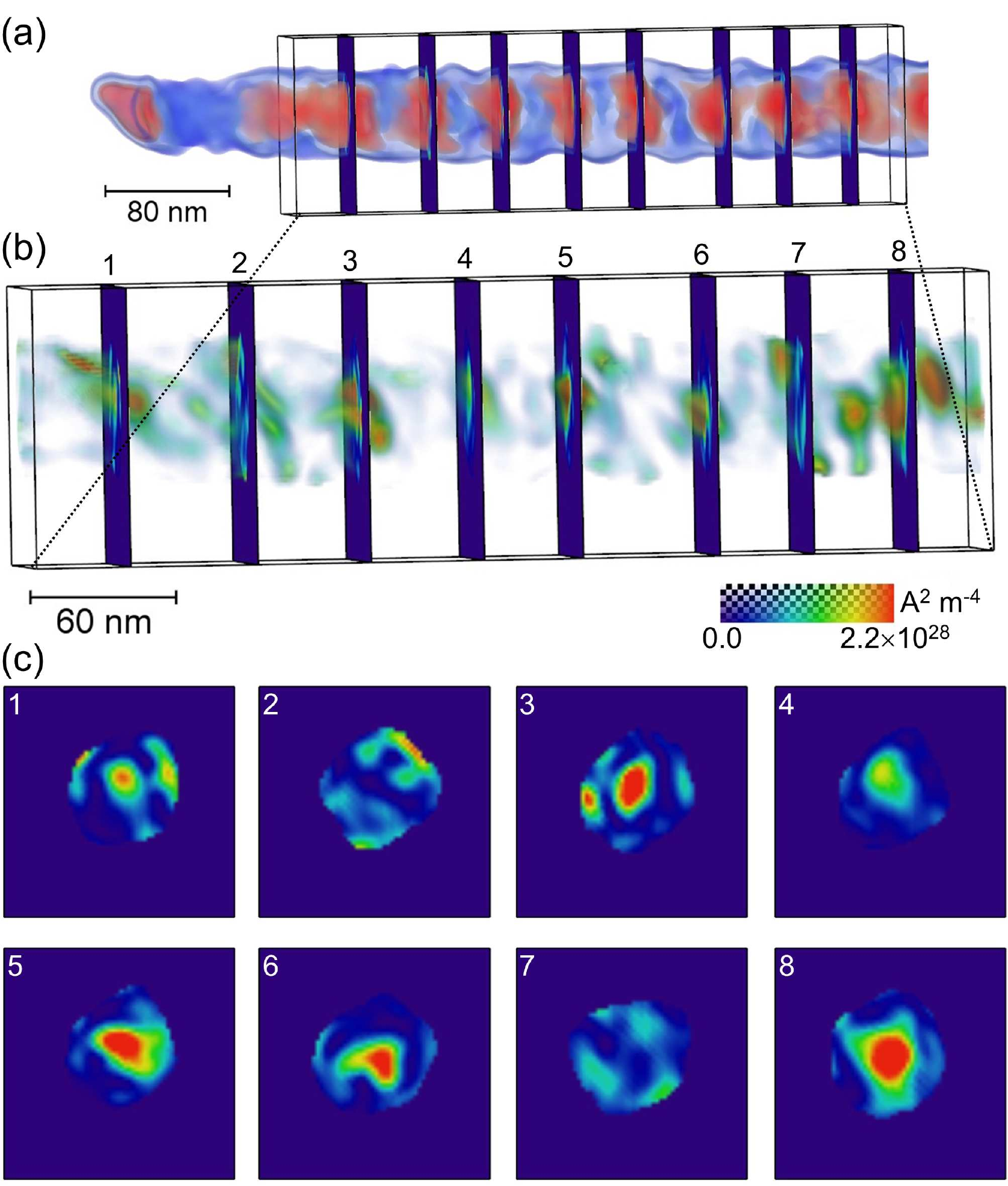}

\caption{\label{fig:Current-exchange-density}Magnetization current exchange
density. (a) \textcolor{black}{3D mean inner potential tomogram of
the Co/Cu nanowire with bounding box indicating the region where the
current exchange density is computed. (b) Volume rendering of the
current exchange density. (c) Slices through Co disks 1-8, in which
the Co disks with a vortex state (1, 3-6, and 8) exhibit a localized
maximum in the current exchange density.}}
\end{figure}

In the following we elaborate on the relation of other important magnetic
quantities to the $\mathbf{B}$-field, in order to comprehensively
characterize the nanomagnetic properties of the NW. We first note
that the conservative (longitudinal) and solenoidal (transverse) part
of the Helmholtz decomposition of the magnetization (with scalar magnetic
potential $\mathit{\Phi}$ and vector potential $\mathbf{A}$)
\begin{equation}
\mathbf{M}=\underset{-\mathbf{H}}{\underbrace{\nabla\mathit{\Phi}}}+\underset{\mu_{0}^{-1}\mathbf{B}}{\underbrace{\mu_{0}^{-1}\nabla\times\mathbf{A}}}
\end{equation}
directly correspond to the magnetic field and magnetic induction.
As holographic VFET reconstructions cannot reconstruct the conservative
part (i.e., $\mathbf{H}$, which is typically large at boundaries,
interfaces but also in N\'eel textures), the magnetization $\mathbf{M}$
cannot be unambiguously retrieved from VFET without additional knowledge
about the magnetism of the sample. In order to elaborate on this crucial
point, we first compute the total current density $\mathbf{j}=\mu_{0}^{-1}\nabla\times\mathbf{B}$,
i.e., the vorticity of the magnetic induction. \textcolor{black}{If
we decompose $\mathbf{j}$ into free and magnetization (or bound)
currents (}$\mathbf{j}=\mathbf{j_{f}}+\mathbf{j_{b}}$\textcolor{black}{)
and take into account that the former can be neglected in the magnetostatic
limit, we obtain the magnetization current from}
\begin{equation}
\mathbf{j_{b}}=\nabla\times\mathbf{M}=\mu_{0}^{-1}\nabla\times\mathbf{B}\,,\label{eq:bound currents}
\end{equation}
which appears in various energy terms of the micromagnetic free energy.
In our case, the three most important micromagnetic energy contributions
determining the remanent state in the stacked NW are the exchange,
demagnetizing field, and crystallographic anisotropy energy 
\begin{equation}
E_{\text{tot}}\left[\mathbf{M}\right]=E_{\text{ex}}\left[\mathbf{M}\right]+E_{\text{d}}\left[\mathbf{M}\right]+E_{\text{a}}\left[\mathbf{M}\right]\,.\label{eq:EnergyFunctional}
\end{equation}
The magnetization current contributes among others to the exchange
energy (see \textcolor{black}{Supplementary }Note~6 for detailed
expressions of $E_{\text{d}}$ and $E_{\text{a}}$)
\begin{align}
E_{\text{ex}}\left[\mathbf{M}\right] & =\frac{A}{M_{\mathrm{s}}^{2}}\int\left(\left(\nabla\cdot\mathbf{M}\right)^{2}+\left|\nabla\times\mathbf{M}\right|^{2}\right)\mathrm{d}V\label{eq:ExEnergy}\\
 & =\frac{A}{M_{\mathrm{s}}^{2}}\int\left(\left(\rho_{\mathrm{m}}\right)^{2}+\left|\mathbf{j_{b}}\right|^{2}\right)\mathrm{d}V\nonumber 
\end{align}
with exchange stiffness $A$, and saturation magnetization $M_{\mathrm{s}}$.
Therein, the first term denotes the magnetic charge contribution ($\rho_{\mathrm{m}}=-\nabla\cdot\mathbf{M}=\Delta\mathit{\Phi}$),
and additional terms describing surface contributions not written
out here (see Methods\textcolor{black}{{} section and Supplementary
}Note~6 for further details). Consequently, exchange energy minimization
favours suppression of both magnetic charge and magnetization current
in the volume. The 3D distribution of the magnetization current exchange
density $\left|\mathbf{j_{b}}\right|^{2}$ in the Co/Cu NW is computed
from the reconstructed $\mathbf{\boldsymbol{B}}$-field using Eq.~\ref{eq:bound currents},
and correlated with the MIP tomogram (Fig.~\ref{fig:Current-exchange-density}a).
As a result, the current exchange density is effectively minimized
in the homogeneously magnetized parts (slices 2,7 in Figs.~\ref{fig:Current-exchange-density}b,c).
In contrast, it is significantly larger in the vortex regions (slices
1,3-6,8 in Figs.~\ref{fig:Current-exchange-density}b,c), which is,
of course, compensated by the decrease in demagnetizing field and
energy in the total energy functional.

\subsection*{Comparison with micromagnetic simulations}

The above considerations now open a way to derive $\mathbf{M}$ from
$\mathbf{B}$ by incorporating micromagnetic modelling. At the example
of the symmetric magnetic exchange energy (Eq. (\ref{eq:ExEnergy}),
see Supplementary Note~6 for the other energy contributions), we
see that the micromagnetic energy functional can be reduced to a functional
of a scalar field (i.e., the magnetostatic potential $\mathit{\Phi}$),
$E_{\text{tot}}\left[\mathit{\mathbf{M}}\right]\rightarrow E_{\text{tot}}\left[\mathit{\Phi}\right]$,
if $\mathbf{B}$ (and hence $\mathbf{j_{b}}$) is known (from the
experiment). This significantly reduces the degrees of freedom and
hence the complexity of the micromagnetic minimization problem of
$E_{\text{tot}}$, which can be exploited in various ways for retrieving
$\mathbf{M}$. First, micromagnetic simulations can be used to compute
$\mathbf{B}$ from a given $\mathbf{M}$ and iterate over different
magnetization configurations until agreement with the reconstructed
data is reached. We use this approach below to obtain information
about the exchange stiffness and magnetocrystalline anisotropy in
the stacked NW. Second, micromagnetic simulation can be adapted such
to directly minimize the energy as a function of the scalar field
$\mathit{\Phi}$. Such adapted micromagnetic schemes, which even allow
to analytically derive the total magnetization $\mathbf{M}$ for highly
symmetric magnetic configurations, are further discussed in Supplementary
Note~6.

\begin{figure*}[t]
\includegraphics[width=1\textwidth]{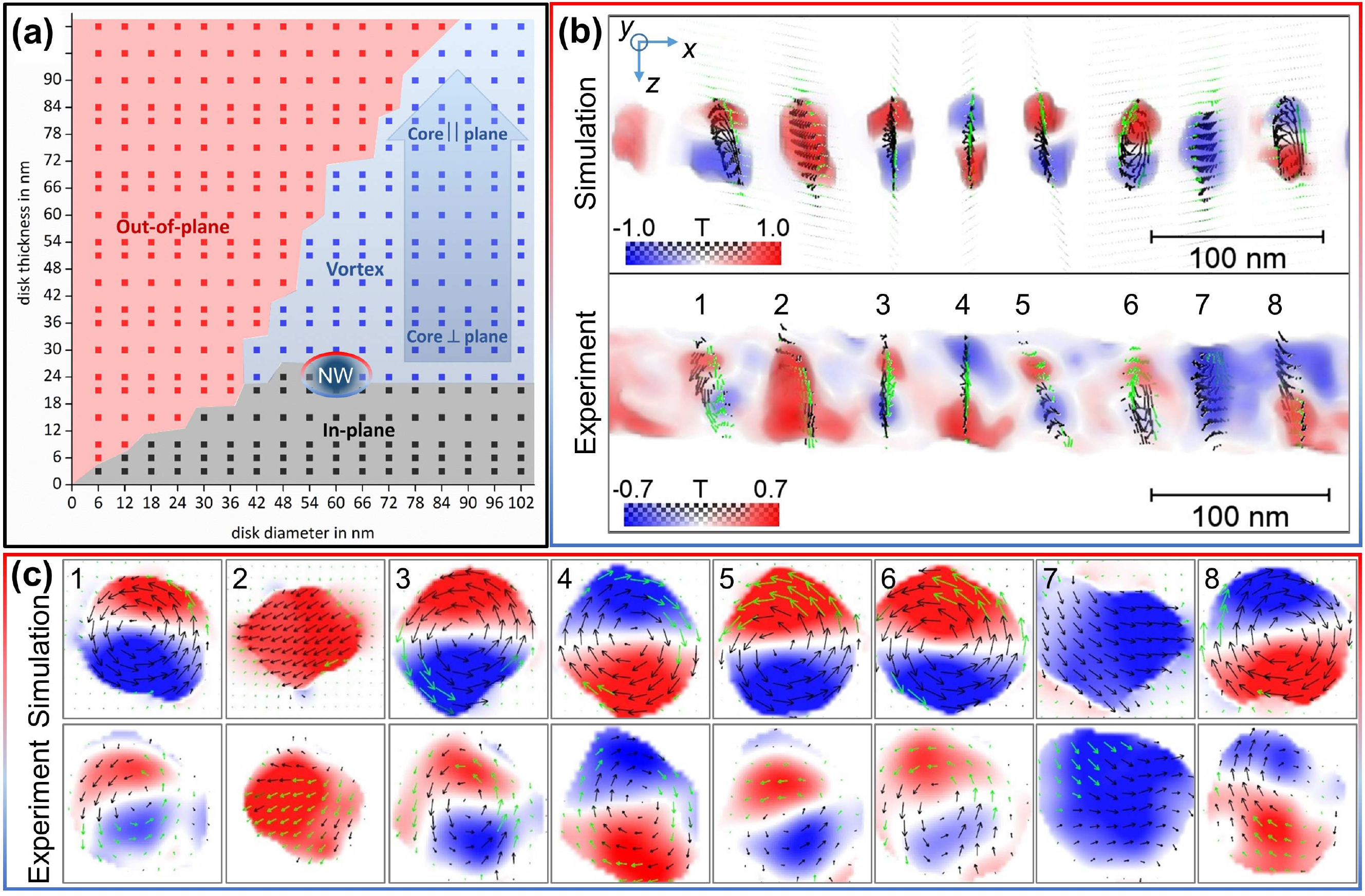}

\caption{\textcolor{red}{\label{fig: Micromag}}\textcolor{black}{Micromagnetic
simulations. (a) Simulated magnetic phase diagram of regular Co disks/cylinders
obtained by variation of diameter and thickness. Each point (square)
corresponds to the resulting magnetic state obtained from one micromagnetic
simulation. The region of the investigated nanowire (NW) is highlighted
around $60$\,nm diameter and $25$\,nm thickness. (b) Comparison
of the $B_{y}$ component between the simulated and experimental Co/Cu
NW. In the simulation, the irregular morphology of the Co disks obtained
from the mean inner potential tomogram was taking into account. Note
the different blue-red color-coding range between simulation and experiment
according to the color bars. (c) }Slices through the Co disks 1-8
as indicated in (b) superimposed by an arrow plot indicating the magnetic
state. The opposite out-of-plane directions are distinguished by black
and green as also shown in the arrowplots in (b).}
\end{figure*}
In the light of the above discussion, we now employ micromagnetic
simulations to gain more insight into the complex magnetic structure
of the Co/Cu NW (see Methods section and Supplementary Note~7 for
details). In order to foresee which kind of magnetic remanent states
could be obtained in such layers, we first simulated the remanent
state phase diagram of a single Co disk as a function of thickness
and diameter (Fig. \ref{fig: Micromag}a). To mimic the experimental
conditions, remanent states were calculated after saturation with
1\,T close to the wire axis. Simulations were performed with the
magnetic parameters of Ref. \citep{Reyes2016} and a tilt angle of
the layers of about 10${^\circ}$ with respect to the wire axis. Depending on
the ratio of thickness and diameter, the remanent states of a single
layer can be out-of-plane, in-plane or vortices with the core either
being out-of-plane or tilted with respect to the normal of the layers.
The Co disks of the investigated NW are at the boundary of the in-plane
and vortex configurations, which agrees well with the coexistence
of both configurations in the experiment. Additional contributions
such as \textcolor{black}{dipolar coupling between the discs, multiple
grains including magnetocrystalline and surface anisotropies, coupling
to defects, irregular disc shapes and chemical gradients further influence}
the magnetic state of a real Co disc. In particular the contribution
of the crystallographic anisotropy is very complex in the Co (alloyed
with 15\% Cu) disks including randomly oriented nanoscaled grains
of predominant fcc symmetry (i.e., cubic anisotropy). Using TEM, the
nano-crystallinity can be observed by local diffraction contrast of
the nanometer-sized grains when they are oriented in low-index zone
axis with respect to the electron beam direction. This is visible
in the bright-field TEM image (Fig. \ref{fig:3D-reconstruction}a)
and also in the original electron holograms, from which the tomograms
are reconstructed (Supplementary Note~8). Therefore, to go deeper
in the analysis of the magnetic properties of the layers, we performed
simulations of the eight Co layers shown in Figs.~\ref{fig: Micromag}b,c.
Within the simulations the magnetization amplitude \textcolor{black}{$M_{\mathrm{s}}$},
exchange constant $A$ and crystalline anisotropy \textcolor{black}{$H_{\mathrm{K}}$}
of each Co layer were changed until the best fit to the experiment
was achieved. A crucial step is to implement the 3D morphology of
the layers extracted from the MIP tomogram into the micromagnetic
simulations (see Methods section and Supplementary Note~7) in order
to take into account the geometrical symmetry breaking and inhomogeneity
of the Co layers in the calculations. This valuable input significantly
reduces the numbers of unknown parameters to reproduce the magnetic
configurations in the simulations, and allowed to extract magnetic
parameters for all individual layers within a range for $M_{\mathrm{s}}$
$\left(1000-1200\right)\,\mathrm{Am^{-1}}$, $A$ $\left(12-20\right)10^{-12}\,\mathrm{Jm^{-1}}$,
and $H_{\mathrm{K}}$ $\left(50-130\right)10^{3}\mathrm{\mathrm{Jm^{-3}}}$
(see Supplementary Table 2 for values of each layer). The variation
of these intrinsic parameters from one layer to the other reflects
the large impact of the individual disc shapes in the formation of
the magnetic configurations. Note, however, the magnetization is found
to be between 20 and 30\%\textcolor{black}{{} overestimated in our simulations
(Fig. \ref{fig: Micromag}c), which indicates that the magnetization
amplitude and most probably the exchange constant for each layer can
be decreased even further. Such low values of magnetic constants due
to the Cu impurities in the Co layers were also observed in a recent
work \citep{Ogrodnik2019} on electrodeposited Co/Cu multilayers grown
in a single bath, especially when decreasing the thickness of the
layer.}

\section*{Discussion}

We have demonstrated the successful 3D reconstruction of both the
magnetic induction vector field and 3D chemical composition of a complex
real nanomaterial with sub-$10\,\text{nm}$ spatial resolution using
holographic VFET. Crucial steps are the semi-automated holographic
acquisition of dual-axis tilt series within a tilt range of 280${^\circ}$,
holographic phase reconstruction, precise image alignment, separation
of electric and magnetic phase shift, the tomographic reconstruction
of \textcolor{black}{all three $\mathbf{B}$-field} components exploiting
the constraint $\nabla\cdot\mathbf{B}=0$. Moreover, we elaborated
on the extraction of magnetic properties such as the solenoidal magnetic
exchange energy from the tomographic data. \textcolor{black}{The results
obtained from a multilayered Co/Cu NW paint a complex picture of the
3D magnetization behaviour, for example, the coexistence of vortex
and in-plane magnetized states. This allows to set up a micromagnetic
model including the exact geometry of the Co nanodisks that matches
the experimental $\mathbf{B}$-field tomogram, from which magnetic
parameters of individual layers can be derived. Indeed, micromagnetic
simulations of nano-objects are generally performed considering \textquotedblleft ideal\textquotedblright{}
systems, which can lead to false predictions of the magnetic states.
Avoiding problems of geometrical uncertainties as well as providing
additional data in terms of 3D $\mathbf{B}$-field distribution for
optimizing micromagnetic simulation is therefore a great advance for
a precise analysis of the magnetic properties of nano-objects. We
anticipate further improvements by including additional tilt series
(e.g., uitilzing improved 3-tilt axis tomography holders), increased
}signal-to-noise ratio (SNR)\textcolor{black}{{} at long exposure times
using automated feedback of the microscope \citep{Gatel2018}, improved
vector field reconstruction schemes and adapted micromagnetic modeling
of the magnetostatic potential, explicitly exploiting the a priori
knowledge of the $\mathbf{B}$-field. The technique holds large potential
for revealing complex 3D nanomagnetization patterns, e.g., in chiral
magnets, nanomagnets (e.g, NWs) and frustrated magnets, currently
not possible with other methods at the required spatial resolution.}

\section*{Methods}

\subsection*{Nanowire Synthesis}

NWs are grown via template-based electrodeposition technique. The
thereby used template was a commercial polycarbonat membrane (Maine
Manufacturing, LLC). NWs with diameter ranging from 55 to 90\,nm
are obtained by electrodeposition in pulse mode from a single bath
containing both Co and Cu ions. The deposition potentials for Co and
Cu are $-1\,\text{V}$ and $-0.3\,\text{V}$, respectively. Lower
deposition potential of the Co leads to about 15\% of Cu impurities
inside the Co layers \citep{Reyes2016}. The duration of the deposition
potential pulses was adjusted to $1\,\text{s}$ for Co and $10\,\text{s}$
for Cu, to achieve nominal layer thicknesses of $25\,\text{nm}$ Co
and $15\,\text{nm}$ Cu. In order to be transferred onto a holey carbon
grid for TEM experiments, the membrane surrounding the wires was removed
by immersion of the sample in dichloromethane. All details about the
growth and cleaning processes are given in Ref. \citep{Reyes2016}
and references therein.\textcolor{black}{{} Prior to the tomography
experiments, the NWs were saturated in a 1T magnetic field oriented
roughly at 10${^\circ}$ from the wire axis direction.}

\subsection*{Acquisition of holographic tilt series}

The holographic tilt series was recorded at the FEI Titan 80-300 Holography
Special Berlin TEM instrument, in image-corrected Lorentz-Mode (conventional
objective lens turned off) operated at 300 kV. We employed the two
electron bisprism setup \citep{Harada2004} that was adjusted by an
upper biprism voltage of $35\,\text{V}$, a lower biprism voltage
of $100\,\text{V}$, and an intermediate X-lens (extra lens between
the two biprisms) excitation of -0.36. For acquisition of electron
holgrams, a 2k by 2k slow scan CCD camera (Gatan Ultrascan 1000 P)
was used. The high tilt angles, and the manual in-plane rotation of
the sample in between the two tilt series (one to reconstruct $B_{x}$
and one to reconstruct $B_{y}$), were achieved by means of a dual-axis
tomography holder (Model 2040 of E. A. Fischone Instruments, Inc.).
The acquisition process was performed semi-automatically with an in-house
developed software package for an efficient collection of holographic
tilt series consisting of object and object-free empty hologram \citep{Wolf2010}.
The latter is required for correction purposes as explained below.
For the first tilt series, the specimen was rotated in-plane, such
that an angle between NW axis and tilt axis of $+44.5{^\circ}$ was
obtained. For the second tilt series, the specimen was rotated further
in-plane by $91{^\circ}$ (ideal is $90{^\circ}$) resulting in an
angle between NW axis and tilt axis of -$46.5{^\circ}$. After turning
the sample upside down ex-situ,  two more holographic tilt series
were recorded that represent the flipped version of the first two
tilt series. The tilt range of each tilt series was from $-69{^\circ}$ to
$+66{^\circ}$ in $3{^\circ}$steps. The electon holograms had an
average contrast of $14\%$ in vacuum, a fringe spacing of $3.3\thinspace\text{nm}$,
a field of view of $1\thinspace\text{\ensuremath{\mu}m}$, and a pixel
size of $0.48\,\text{nm}$.

\subsection*{Reconstruction and processing of projection data}

The elaborate image data treatment was mainly accomplished using in-house
developed scripts and software plugins for Gatan Microscopy Suite
(GMS). Amplitude and phase images were reconstructed by filtering
out one sideband of the Fourier transform of the electron hologram,
e.g., described in Ref.~\citep{Lehmann2002}). Likewise, amplitude
and phase images were reconstructed from empty holograms for correction
of imaging artifacts, such as distortions of camera and projective
lenses. In detail, we used a soft numerical aperture with a full width
at half maximum of $1/9\,\text{nm}^{-1}$, and zero-damping the signal
at $1/4.5\,\text{nm}^{-1}$. Accordingly, the resolution (reconstructed
pixel size) of the image wave after inverse Fourier transformation
(FT) of the cut sideband is in the range from $4.5\thinspace\text{nm}$
to $9\,\text{nm}$ depending on the SNR. After holographic reconstruction
of all four tilt series, phase images were unwrapped with automatic
phase unwrapping algorithms (Goldstein or Flynn) \citep{Ghiglia1998}.
Possible artifacts after application of these algorithms at regions,
where the phase signal is too noisy or undersampled, are corrected
by manual treatment of using preknowledge of the phase shift (e.g.
from adjacent projections) \citep{Lubk2014}. All four phase tilt
series were prealigned separately by cross-correlation to correct
for coarse displacements between successive projections. Furthermore,
those two tilt series, where the sample was flipped before acquisition,
were flipped back numerically yielding for each tilt angle a pair
of phase images with equal electric but opposite magnetic phase shift.
Then, each pair of phase images was aligned by applying a linear affine
transformation on the \textquotedblleft flipped\textquotedblright{}
phase image, which considers displacements, rotation, and direction-dependent
magnification changes \citep{Wolf2015}. After this step, the separation
of electric and magnetic phase shifts was performed for these pairs
of phase images as illustrated in Fig.~\ref{fig:Principle-of-EHVT},
steps (5) and (6). To finally employ the fine alignment (i.e., the
accurate determination of the tilt axis and correction for subpixel
displacement), we used a self-implemented center-of-mass method for
correction of displacements perpendicular to the tilt axis and the
common line approach \citep{Penczek1996} for the correction of displacements
parallel to the tilt axis. We applied these procedures initially on
the electric phase tilt series, which has a higher SNR than the magnetic
one, and used the thereby determined displacements for alignment of
magnetic phase tilt series. Before computation of the derivatives,
the magnetic phase images were smoothed slightly by a non-linear anisotropic
diffusion (NAD)\citep{Frangakis2001} filter using the Avizo software
package (ThermoFisher Scientific Company). Following Eq. \ref{eq:PhaseGrad_vs_B},
we calculated the projected $B_{x}$- and $B_{y}$- components from
the derivatives of the magnetic phase images in $y$- and $x$- direction.

\subsection*{Tomographic reconstruction}

The tomographic reconstruction of electrostatic potential, $B_{x}$-
and $B_{y}$- component from their aligned tilt series (projections)
was conducted with the weighted simultaneous iterative reconstruction
technique (W-SIRT) \citep{Wolf2014}. Since the W-SIRT involves a
weighted (instead of a simple) back-projection for each iteration
step, it convergences faster than a conventional SIRT algorithm. The
number of iterations was determined to five, by visually inspecting
the trade-off between spatial resolution and noise. In addition, we
reduced the missing wedge artifacts in the tomograms at low spatial
frequencies by a finite support approach, explained in Ref. \citep{Wolf2018}.

\subsection*{Calculation of the third magnetic B-field component}

In order to determine the third $\mathbf{B}$-field component $B_{z}$
not directly obtainable from one of the two tilt series around $x$
or $y$, two strategies may be applied. First, it is possible to solve
third Maxwell's law $\text{div}\mathbf{B}=0$ with appropriate boundary
conditions on the surface of the reconstruction volume. Second, it
is possible to compute the projected component of the $\mathbf{B}$-field
perpendicular to the tilt axis in one tilt series (say around $x$),
which is a mixture of the $y$ and $z$ component in the non-rotated
laboratory frame and substitute the $y$ component with the reconstruction
from the second tilt series around $y$. The second approach may be
implemented in both a field- and vector potential-based reconstruction
scheme (automatically implying $\text{div}\mathbf{B}=0$) \citep{Phatak(2008)}.
Indeed, the two strategies to determine $B_{z}$ are identical as
demonstrated in the Supplementary Note~1. 

The most straight-forward way to calculate $B_{z}$ from Eq. \ref{eq:Gauss's law for magnetism}
is by integration along the $z$-coordinate, that is,

\begin{equation}
B_{z}\left(x,y,z\right)=-\int_{-\infty}^{z}\left(\partial_{x}B_{x}\left(x,y,z'\right)+\partial_{y}B_{y}\left(x,y,z'\right)\right)\mathbf{\mathrm{d}}z'.
\end{equation}

However, to reduce accumulation of errors while integrating along
$z$, we employed periodic boundary conditions for solving this differential
equation, which reads in Fourier space 
\begin{equation}
B_{z}\left(x,y,z\right)=\mathfrak{\mathcal{F}}^{-1}\left\{ -\frac{k_{x}}{k_{z}}\mathfrak{\mathcal{F}}\left\{ B_{x}\right\} -\frac{k_{x}}{k_{z}}\mathfrak{\mathcal{F}}\left\{ B_{y}\right\} \right\} .
\end{equation}
Here $\mathcal{F}\left\{ \right\} $ and $\mathcal{F}^{-1}\left\{ \right\} $
denote the forward and inverse 3D FT, and $k_{x.},k_{y},k_{z}$ the
reciprocal coordinates in 3D Fourier space. The zero frequency component
(integration constant) was fixed by setting the average of $B_{z}$
to zero on the boundary of the reconstruction volume.

\subsection*{Micromagnetics}

Micromagnetic simulations of the remanent magnetic states were perform\textcolor{black}{ed
with Object Oriented Micromagnetic Framework (OOMMF) software package
\citep{Donahue_1999} solving the} non-linear minimization problem
(non-linear constraint $\left|\mathbf{M}\right|=M_{\mathrm{s}}$)
of the energy functional (Eq. \ref{eq:EnergyFunctional}) containing
the exchange, demagnetizing field and crystalline anisotropy energy
as the main contributions in our \textcolor{black}{case. The details
of the single disc simulations are given in the Supplementry Note~7.
The 3D morphology of the Co/Cu layers is revealed from MIP tomogram
by assigning each voxel to a different material region. The NW's 3D
shape is segmented by a MIP threshold of $11.0\,\text{V}$ to distinguish
between NW and vacuum, whereas a MIP threshold of $18.5\,\text{V}$
was employed to distinguish between Co ($>18.5\,\text{V}$) and Cu
($\leq18.5\,\text{V}$). Then, the 3D data array is interpolated to
a grid with voxels of 2 x 2 x 2\,$\mathrm{nm}^{3}$ (grid dimension:
960 x 160 x 200 $\mathrm{nm}^{3}$). Each Co layer is attributed to
an individual set of magnetic parameters such as magnetization $M_{\mathrm{s}}$,
exchange constant $A$ and uniaxial crystal anisotropy $H_{\mathrm{K}}$
(see Supplementary Table~2). Here, the assumption of uniaxial anisotropy
relies on the polycrystalline nature of the layers, which may be approximated
by an average anisotropy in a first order approximation. The magnetic
induction $\mathbf{B}$ in each layer is obtained summing the calculated
magnetization components and the demagnetization field $\mathbf{H}$.}

\section*{Data availability}

The data that support the findings of this study are available from
the corresponding author upon reasonable request.

\section*{Acknowledgement}

DW, AL, SS, and JK have received funding from the European Research
Council (ERC) under the Horizon 2020 research and innovation programme
of the European Union (grant agreement No 715620). The authors further
express their gratitude to M. Lehmann (TU Berlin) for providing access
to the FEI Titan 80-300 Holography Special Berlin.

\section*{Author contributions}

D.W., S.S., A.L., C.G., B.W. and T.N. performed the experiments. D.W.
aligned, reconstructed, and evaluated the 3D data. T.W., N.B. and
D.R. fabricated the samples. N.B. carried out the micromagnetic simulations.
A.L. and J.K. contributed to the theoretical foundation of VFET and
derivation of magnetic properties. D.W., and A.L. wrote the manuscript.
N.B. and C.G. contributed to manuscript writing. E.S. and B.B. helped
with the interpretation of the data and commented on the manuscript.

\section*{Additional information}

\subsection*{Competing interests}

The authors declare no competing interests.

\section*{References}

\begin{thebibliography}{10}

\bibitem{Fernandez-Pacheco2017}
Fernandez-Pacheco, A., Streubel, R., Fruchart, O., Hertel, R., Fischer, P., and
  Cowburn, R.~P.
\newblock {\em Nat Commun}{ \bf 8}, 15756 (2017).

\bibitem{Parkin2008}
Parkin, S.~S., Hayashi, M., and Thomas, L.
\newblock {\em Science}{ \bf 320}, 190--194 (2008).

\bibitem{karnaushenko_self-assembled_2015}
Karnaushenko, D., Karnaushenko, D.~D., Makarov, D., Baunack, S., Sch\"afer, R.,
  and Schmidt, O.~G.
\newblock {\em Advanced Materials}{ \bf 27}, 6582--6589 (2015).

\bibitem{hoffmann_opportunities_2015}
Hoffmann, A. and Bader, S.~D.
\newblock {\em Physical Review Applied}{ \bf 4}, 047001 (2015).

\bibitem{fruchart_bloch-point-mediated_2018}
Fruchart, O., Wartelle, A., Trapp, B., Stano, M., Thirion, C., Bochmann, S.,
  Bachmann, J., Foerster, M., Aballe, L., Mentes, O., Locatelli, A., Sala, A.,
  Cagnon, L., and Toussaint, J.-C.
\newblock {\em arXiv:1806.10918 [cond-mat]}{ \bf } (2018).

\bibitem{Biziere2013}
Biziere, N., Gatel, C., Lassalle-Balier, R., Clochard, M.~C., Wegrowe, J.~E.,
  and Snoeck, E.
\newblock {\em Nano Lett}{ \bf 13}, 2053--2057 (2013).

\bibitem{Bogdanov1989}
Bogdanov, A.~N. and Yablonskii, D.~A.
\newblock {\em Zh. Eksp. Teor. Fiz.}{ \bf 95}, 178--182 (1989).

\bibitem{Rybakov2013}
Rybakov, F.~N., Borisov, A.~B., and Bogdanov, A.~N.
\newblock {\em Physical Review B - Condensed Matter and Materials Physics}{ \bf
  87}, 094424 (2013).

\bibitem{Rybakov2016}
Rybakov, F.~N., Borisov, A.~B., Bl{\"{u}}gel, S., and Kiselev, N.~S.
\newblock {\em New Journal of Physics}{ \bf 18}, 045002 (2016).

\bibitem{Zheng2018}
Zheng, F., Rybakov, F.~N., Borisov, A.~B., Song, D., Wang, S., Li, Z.-A., Du,
  H., Kiselev, N.~S., Caron, J., Kovacs, A., Tian, M., Zhang, Y., Bl{\"{u}}gel,
  S., and Dunin-Borkowski, R.~E.
\newblock {\em Nature Nanotechnology}{ \bf 13}, 451--455 (2018).

\bibitem{Kravchuk2018}
Kravchuk, V.~P., Sheka, D.~D., K\'akay, A., Volkov, O.~M., R\"o\ss{}ler, U.~K.,
  van~den Brink, J., Makarov, D., and Gaididei, Y.
\newblock {\em Phys. Rev. Lett.}{ \bf 120}, 067201 (2018).

\bibitem{Nisoli2013}
Nisoli, C., Moessner, R., and Schiffer, P.
\newblock {\em Rev. Mod. Phys.}{ \bf 85}, 1473--1490 (2013).

\bibitem{Streubel2015}
Streubel, R., Kronast, F., Fischer, P., Parkinson, D., Schmidt, O.~G., and
  Makarov, D.
\newblock {\em Nature Communications}{ \bf 6}, 7612 (2015).

\bibitem{suzuki_three-dimensional_2018}
Suzuki, M., Kim, K.-J., Kim, S., Yoshikawa, H., Tono, T., Yamada, K.~T.,
  Taniguchi, T., Mizuno, H., Oda, K., Ishibashi, M., Hirata, Y., Li, T.,
  Tsukamoto, A., Chiba, D., and Ono, T.
\newblock {\em Applied Physics Express}{ \bf 11}, 036601 (2018).

\bibitem{donnelly_tomographic_2018}
Donnelly, C., Gliga, S., Scagnoli, V., Holler, M., Raabe, J., Heyderman, L.~J.,
  and {Manuel Guizar-Sicairos}.
\newblock {\em New Journal of Physics}{ \bf 20}, 083009 (2018).

\bibitem{Lai1994}
Lai, G., Hirayama, T., Fukuhara, A., Ishizuka, K., Tanji, T., and Tonomura, A.
\newblock {\em Journal of Applied Physics}{ \bf 75}, 4593--4598 (1994).

\bibitem{Lubk(2014)d}
Lubk, A., Wolf, D., Simon, P., Wang, C., Sturm, S., and Felser, C.
\newblock {\em Applied Physics Letters}{ \bf 105}, 173110 (2014).

\bibitem{Wolf2015}
Wolf, D., Rodriguez, L.~A., Beche, A., Javon, E., Serrano, L., Magen, C.,
  Gatel, C., Lubk, A., Lichte, H., Bals, S., Van~Tendeloo, G.,
  Fernandez-Pacheco, A., De~Teresa, J.~M., and Snoeck, E.
\newblock {\em Chem Mater}{ \bf 27}, 6771--6778 (2015).

\bibitem{Simon2016}
Simon, P., Wolf, D., Wang, C., Levin, A.~A., Lubk, A., Sturm, S., Lichte, H.,
  Fecher, G.~H., and Felser, C.
\newblock {\em Nano Lett}{ \bf 16}, 114--120 (2016).

\bibitem{Lichte2008a}
Lichte, H.
\newblock {\em Ultramicroscopy}{ \bf 108}, 256--262 (2008).

\bibitem{Roder(2014)a}
R\"oder, F., Lubk, A., Wolf, D., and Niermann, T.
\newblock {\em Ultramicroscopy}{ \bf 144}, 32--42 (2014).

\bibitem{Phatak(2010)}
Phatak, C., Petford-Long, A.~K., and De~Graef, M.
\newblock {\em Phys. Rev. Lett.}{ \bf 104}, 253901 (2010).

\bibitem{Tanigaki(2015)}
Tanigaki, T., Takahashi, Y., Shimakura, T., Akashi, T., Tsuneta, R., Sugawara,
  A., and Shindo, D.
\newblock {\em Nano Letters}{ \bf 15}, 1309--1314 (2015).

\bibitem{Caron2018}
Caron, J.
\newblock {\em Model-based reconstruction of magnetisation distributions in
  nanostructures from electron optical phase images}, volume 177 of {\em Key
  Technologies}.
\newblock Forschungszentrum J\"ulich GmbH Zentralbibliothek, Verlag J\"ulich,
  (2018).

\bibitem{Mohan2018}
Mohan, K.~A., Kc, P., Phatak, C., De~Graef, M., and Bouman, C.~A.
\newblock {\em IEEE Transactions on Computational Imaging}{ \bf 4}, 432--446
  (2018).

\bibitem{Aharonov1959}
Aharonov, Y. and Bohm, D.
\newblock {\em Physical Review}{ \bf 115}, 485--491 (1959).

\bibitem{Teague(1983)}
Teague, M.
\newblock {\em Journal of the Optical Society of America}{ \bf 73}, 1434--1441
  (1983).

\bibitem{Lubk2018a}
Lubk, A.
\newblock In {\em Advances in Imaging and Electron Physics}, volume 206.
  (2018).

\bibitem{Kasama2011}
Kasama, T., Dunin-Borkowski, R., and Beleggia, M.
\newblock {\em Electron Holography of Magnetic Materials}.
\newblock InTech (2011).

\bibitem{Midgley2003}
Midgley, P.~A. and Weyland, M.
\newblock {\em Ultramicroscopy}{ \bf 96}, 413--431 (2003).

\bibitem{Reyes2016}
Reyes, D., Biziere, N., Warot-Fonrose, B., Wade, T., and Gatel, C.
\newblock {\em Nano Lett}{ \bf 16}, 1230--1236 (2016).

\bibitem{Heel2005}
van Heel, M. and Schatz, M.
\newblock {\em J Struct Biol}{ \bf 151}, 250--262 (2005).

\bibitem{Ogrodnik2019}
Ogrodnik, P., Vetro, F.~A., Frankowski, M., Checinski, J., Stobiecki, T.,
  Barnas, J., and Ansermet, J.-P.
\newblock {\em Journal of Physics D: Applied Physics}{ \bf 52}, 065002 (2019).

\bibitem{Gatel2018}
Gatel, C., Dupuy, J., Houdellier, F., and Hytch, M.~J.
\newblock {\em Applied Physics Letters}{ \bf 113}, 133102 (2018).

\bibitem{Harada2004}
Harada, K., Tonomura, A., Togawa, Y., Akashi, T., and Matsuda, T.
\newblock {\em Applied Physics Letters}{ \bf 84}, 3229--3231 (2004).

\bibitem{Wolf2010}
Wolf, D., Lubk, A., Lichte, H., and Friedrich, H.
\newblock {\em Ultramicroscopy}{ \bf 110}, 390--399 (2010).

\bibitem{Lehmann2002}
Lehmann, M. and Lichte, H.
\newblock {\em Microsc Microanal}{ \bf 8}, 447--466 (2002).

\bibitem{Ghiglia1998}
Ghiglia, D.~C. and Pritt, M.~D.
\newblock {\em Two-dimensional phase unwrapping: Theory, algorithms and
  software}.
\newblock {Wiley {\&} Sons}, New York NY,  (1998).

\bibitem{Lubk2014}
Lubk, A., Wolf, D., Prete, P., Lovergine, N., Niermann, T., Sturm, S., and
  Lichte, H.
\newblock {\em Physical Review B}{ \bf 90}, 125404 (2014).

\bibitem{Penczek1996}
Penczek, P.~A., Zhu, J., and Frank, J.
\newblock {\em Ultramicroscopy}{ \bf 63}, 205--218 (1996).

\bibitem{Frangakis2001}
Frangakis, A.~S. and Hegerl, R.
\newblock {\em J Struct Biol}{ \bf 135}, 239--250 (2001).

\bibitem{Wolf2014}
Wolf, D., Lubk, A., and Lichte, H.
\newblock {\em Ultramicroscopy}{ \bf 136}, 15--25 (2014).

\bibitem{Wolf2018}
Wolf, D., Hubner, R., Niermann, T., Sturm, S., Prete, P., Lovergine, N.,
  Buchner, B., and Lubk, A.
\newblock {\em Nano Lett}{ \bf 18}, 4777--4784 (2018).

\bibitem{Phatak(2008)}
Phatak, C., Beleggia, M., and De~Graef, M.
\newblock {\em Ultramicroscopy}{ \bf 108}, 503--513 (2008).

\bibitem{Donahue_1999}
Donahue, M. J.;~Porter, D.~G.
\newblock {\em OOMMF User's Guide Version 1.0, 1999, Interagency Report NISTIR
  6376}.
\newblock National Institute of Standards and Technology: Gaithersburg, MD.

\end{thebibliography}
\end{document}